\documentclass[twocolumn]{autart}   

\usepackage{curve2e}
\usepackage{graphicx} % Allows including images
\usepackage{booktabs} % Allows the use of \toprule, \midrule and \bottomrule in tables
\usepackage[italian,english]{babel}
\usepackage[T1]{fontenc}
\usepackage[utf8]{inputenc}
\usepackage{subfigure}
\usepackage{multirow}
\usepackage{mathrsfs}
\usepackage{MnSymbol}%
\usepackage{wasysym}%

\def\define{\buildrel \triangle \over =}
\def\reali{{\rm I \! R}}

\begin{document}

\begin{frontmatter}

\title{Steady-state input calculation for achieving a desired steady-state output of a linear systems}% Title, preferably not more 

\thanks[footnoteinfo]{This paper was not presented at any IFAC 
meeting. Corresponding author Raffaele Romagnoli. E-mail: rromagno@ulb.ac.be\\
This work is performed in the framework of the BATWAL project financed by the Walloon region (Belgium).\\
\hspace*{0.14cm} This research has been funded by the Mandats d'Impulsion Scientific "Optimization-free Control of Nonlinear Systems subject to Constraints" of the Fonds de la Recherche Scientifique (FNRS), Ref. F452617F.}

\author[First]{Raffaele Romagnoli} 
\author[First]{Emanuele Garone}

\address[First]{Universit\'e Libre de Bruxelles, B-1050 Bruxelles, Belgium
   (e-mail:{rromagno@ulb.ac.be, egarone}@ulb.ac.be).}

% \begin{keyword}                           % Five to ten keywords,  
% feedforward control; stable inversion; least squares; basis functions               % chosen from the IFAC 
% \end{keyword}                             % keyword list or with the 
                                          % help of the Automatica 
                                          % keyword wizard
 \begin{abstract}                          
 In this note we provide an algorithm for the computation of the steady-state input able to achieve the steady-state output tracking of any desired output signal representable as a rational transfer function. 
\end{abstract}

\end{frontmatter}

\section{Introduction}

Output tracking problem is a well known problem in the literature which has been solved for the tracking of steady-state output signals. One of the methods developed to solve this problem is the internal model principle \cite{francis1976internal}. The existence of the steady-state response needs the asymptotic stability of the considered system. \\ 
Given a particular steady-state behavior of the output, and an asymptotically stable linear systems, it is possible to find the corresponding steady-state input under the hypothesis that the considered system is right invertible \cite{moylan1977stable}. Then a feedforward controller that provides the exact input signal to achieve the output tracking at steady-state can be found. \\  
Despite the computation of the steady state response of a linear system is well known in the literature and can be found in every textbook (e.g. \cite{goodwin2001control}), at the best of the author's knowledge, the inverse procedure has not been developed in the same manner, except for some of the most common and simplest cases. The aim of this paper is to provide an explicit inversion algorithm, for computing the steady-state input for the general case of reference signals defined as fractional trasnfer functions.
\section{Problem Statement}
We consider an Asymptocally Stable Linear Time Invariant SISO System in the form 
\begin{eqnarray}
\dot{x}(t)&=& Ax(t)+Bu(t)\nonumber\\
y(t)&=& Cx(t)+D u(t) \label{sys}
\end{eqnarray}
with $y(\cdot) \in \reali^q$,$u(\cdot) \in \reali^p$ and $x(\cdot)\in \reali^n$ denoting the output, the input, and the state, respectively. 
Its transfer function is denoted by $W(s)=C(sI-A)^{-1}B+D$. Under the hypothesis of right-invertibility \cite{moylan1977stable} the assumption that the system is SISO is without loss of generality as the same  results are directly applicable also to the MIMO case. For this class of systems we will show how to compute input signals ensuring the feedforward asymptotic tracking of the following signals:
\begin{enumerate}
\item $\tilde{y}_d(t)=t^k$ polynomial signal ;
\item $\tilde{y}_d(t)=Y_d\mathrm{sin}(\omega_dt+\psi_d)$ sinusoidal signal ;
\item $\tilde{y}_d(t)=Y_d e^{a_dt}$ with $a_d>0$ exponential signal ;
\item $\tilde{y}_d(t)=e^{a_dt}Y_d\mathrm{sin}(\omega_dt + \psi_d)$ pseudo-periodic signal;
\item $\tilde{y}_d(t)=t^k e^{a_dt} sin(\omega_d t+ \psi_d)$ polynomial pseudo-periodic signal.
\end{enumerate}
\section{Input computation algorithms}
\subsection{Polynomial inputs of order $k$.}
To find the the input $u_s(t)$ such that $\tilde{y}_d(t)=t^k$, we consider the asymptotic output response to the polynomial input $t^k$ that is 
\begin{equation}
\tilde{y}(t)=C_0t^k+C_1t^{k-1}+\cdots+C_k, \label{eq:pol_output}
\end{equation}
where 
\begin{equation}
C_i=\frac{1}{i!}\left[\frac{d^i W(s)}{ds^i}\right]_{s=0}.
\end{equation}
Giving as input the signal
\begin{equation}
u_0(t)=\frac{1}{C_0}t^k,
\end{equation}
the resulting asymptotic output is 
\begin{equation}
\tilde{y}(t)=t^k+ \hat{C}_1 t^{k-1}+\cdots+\hat{C}_k \label{eq:pol_output1}
\end{equation}
where $\hat{C}_i=\frac{C_i}{C_0}$ for $i=1,...,k$. The idea is to find $k$ inputs able to cancel the $\hat{C}_i$ terms of (\ref{eq:pol_output1}). This can be done using the following algorithm
\begin{itemize}
    \item  \textit{Step 0}: $u_0(t)=\frac{t^k}{C_0}$, $\hat{C}_i=\frac{C_i}{C_0}$ for $i=1,...,k$;   
    \item \textit{Step 1:} \textbf{for} $i=1:k$\\  
    \hspace*{1.5cm}     $u_i(t)=-\frac{\hat{C_i}}{C_0}t^{k-i}$;\\
    \hspace*{1.5cm}     \textbf{for} $j=1:k-i$\\
    \hspace*{2cm}     $\hat{C}_{i+1}=\hat{C}_{i+1}-\frac{\hat{C}_i}{C_0}C_j$;\\
    \hspace*{1.5cm}     \textbf{end}\\
    \textbf{end}\\
    \item \textit{Step 2:} $u_s(t)=\sum_{i=0}^k u_i(t)$   
  \end{itemize}
  
\subsection{Sinusoidal signal}
Considering the asymptotic output response of a sinusoidal input of the form $U\mathrm{sin}(\omega_0t+\psi)$
\begin{equation}
\tilde{y}(t)= \vert W(j\omega_0)\vert\ U\mathrm{sin}(\omega_0 t+ (\psi+\mathrm{arg}(G(j\omega_0)))) \label{output:sin}
\end{equation}
and defining as the desired asymptotic response  
\begin{equation}
\tilde{y}_d(t)=Y_d \mathrm{sin}(\omega_d t + \psi_d) \label{des_out:sin}
\end{equation}
the input that generates \eqref{des_out:sin} is $u_s(t)=U sin(\mathrm{\omega_d t + \psi})$ where
\begin{eqnarray}
U&=&\frac{Y_d}{\vert W(j\omega_d)\vert} \label{mod_sin}\\ 
\psi&=&\psi_d - \mathrm{arg}(W(j\omega_d)) \label{phase_sin}
\end{eqnarray}

\subsection{Exponential signal, with positive real exponent $a > 0$} The asymptotic response of an exponential signal of the form $u(t)=U e^{at}$ is
\begin{equation}
\tilde{y}(t)=W(a)U e^{at}.
\end{equation}
Hence, considering the desired output of the form $\tilde{y}_d(t)=Y_d e^{a_dt}$, it can be obtained using $u_s(t)=U e^{a_dt}$ where
\begin{equation}
U=\frac{Y_d}{W(a_d)}
\end{equation}

\subsection{Pseudo-periodic signals} Following the same procedure of the previous cases, we need to find the system response of the following input $u(t)=e^{at}U\mathrm{sin}(\omega t + \psi)$ with a positive real number $a > 0$. To do so, we define the following transfer function
\begin{equation}
H(s)\define \mathcal{L}\left\lbrace w(t)e^{at} \right\rbrace
\end{equation}
where $w(t)=\mathcal{L}^{-1}\left\lbrace W(s)\right\rbrace$, then the output response is
\begin{equation}
\tilde{y}(t)=e^{at}\vert H(j\omega_0)\vert U \mathrm{sin}(\omega_0 t +\psi +\mathrm{arg}(H(j\omega_0))).
\end{equation}
Considering as desired function
\begin{equation}
\tilde{y}_d(t)=e^{a_dt}\vert H(j\omega_d)\vert Y_d \mathrm{sin}(\omega_d t +\psi_d +\mathrm{arg}(H(j\omega_d))),
\end{equation}
the input able to track the desired output is $u_s(t)=e^{a_dt}U\mathrm{sin}(\omega_d t + \psi)$ where $U$ and $\psi$  are computed using (\ref{mod_sin}) and (\ref{phase_sin}).

\subsection{Polynomial pseudo-periodic signals} In this last case all the previous signals are taken into account and the steady state response of an input signal of the form $u(t)=t^k e^{at} sin(\omega t+ \psi)$ with a positive real number $a$ is
\begin{eqnarray}
&&\tilde{y}(t)=e^{at}\left\lbrace \sum_{i=0}^k \left(\begin{array}{c}
k\\
i
\end{array} \right) t^{k-i} \vert H_i(j\omega)\vert \cdot \right. \nonumber\\
&& \hspace*{2.5cm} \left. \begin{array}{c}
\; \\
\;
\end{array}\mathrm{sin}(\omega t + \psi +\mathrm{arg}(H_i(j\omega))) \right\rbrace, \label{poly_pseudo_res}
\end{eqnarray}
where 
\begin{equation}
H_i(s)\define \mathcal{L}\left\lbrace t^i w(t)e^{at} \right\rbrace,
\end{equation}
and $w(t)=\mathcal{L}^{-1}\left\lbrace W(s)\right\rbrace$. If the desired output $\tilde{y}_d(t)$ is expressed as \eqref{poly_pseudo_res}, the proposed algorithm for for polynomial inputs (case 1) can be rearranged taking into account the pseudo-periodic signals to compute $u_s(t)$.

%\bibliographystyle{ieeetr}
%\bibliography{autosam}     

\end{document}